\documentclass[12pt]{article}

\usepackage{amssymb,bbm}
\usepackage{amsfonts}
\usepackage{latexsym}
\usepackage{amsmath}
\usepackage{amsthm}
\usepackage{geometry} 

\def\lddots{\mathinner{\mkern1mu\raise1pt\hbox{.}\mkern2mu  
\raise4pt\hbox{.}\mkern2mu\raise7pt\vbox{\kern7pt\hbox{.}}\mkern1mu}}
\makeatletter
\def\numberbysection{\@addtoreset{equation}{section}
 \def\theequation{\thesection.\arabic{equation}}}
\makeatother  
  
\numberbysection

\newcommand{\be}{\begin{eqnarray}}  
\newcommand{\ee}{\end{eqnarray}} 
\newcommand{\nn}{\nonumber}  
\def\ds{\displaystyle}

\def\bb{\mathbbm}

\def\Z{\bb Z}

\def\a{\alpha}
\def\b{\beta}
\def\d{\delta}
\def\D{\Delta}

\def\l{\lambda}

\def\r{\rho}

\def\t{\theta}

\def\g{\gamma}

\def\pdi #1{\partial_{#1}}

\begin{document}

\vspace*{5.5cm}

\begin{center}
{\LARGE \bf On Calogero-Fran\c{c}oise-type Lax matrices and their dynamical $r$-matrices}

\vspace{1.75cm}

{\large \bf Jean Avan\footnote{e-mail: avan@u-cergy.fr}, Genevi\`eve Rollet\footnote{e-mail: rollet@u-cergy.fr}}

\vspace{15mm}

{\it Laboratoire de Physique Th\'eorique et Mod\'elisation\\
UMR 8089 CNRS, Universit\'e de Cergy-Pontoise, St-Martin II,
2 avenue Adolphe Chauvin, F-95302 Cergy-Pontoise Cedex, France.}
\end{center}

\vspace{1.5cm}

\begin{abstract}
\noindent
New classical integrable systems of Camassa-Holm peakon type are proposed.
They realize the maximal even piecewise-$D_2$ generalization of the Calogero-Fran\c{c}oise
flows, yielding periodic and pseudoperiodic trigonometric/hyperbolic
potentials. The associated  $r$-matrices 
are computed. They are dynamical and depend on both sets $\{p_i\}$ and $\{q_i\}$ of canonical variables. 
\end{abstract}

\vspace{1.5cm}

\setcounter{footnote}{0}
\section{Introduction}

Classical $r$-matrices for the Lax representation of Liouville-integrable systems, depending
on some of the canonical variables (so called ``dynamical $r$-matrices''), have been the subject of a very extensive set
of studies in these past years, starting with the first mention of their relevance~\cite{sts}, and their use in non-ultra local Poisson algebras~\cite{maillet,maillet2}. 

Most studies afterwards
have been concerned with a subset of such matrices, obeying the so-called ``dynamical Yang Baxter algebra''. A paradigmatic
example occurred in the Calogero-Moser $r$-matrix formulation~\cite{AT}, and classification theorems
were extensively developed,  see for instance~\cite{ES}. These matrices depend only on one half of  Poisson-commuting canonical
variables (either momenta or positions), identified in the general
theory~\cite{Fel} as coordinates on the dual of an Abelian Cartan algebra. The
non-abelian situation is discussed in e.g.~\cite{Ping}. 

The above-mentioned associated dynamical classical Yang Baxter equation,
obtained as a sufficient consistency condition for associativity of the Poisson brackets, can then be
related to dynamical quantum algebras such as the Gervais Neveu Felder~(GNF) algebra~\cite{GN,Fel}
or the dynamical reflection algebras~\cite{ArChFr,ANR,Kor}.

More precisely: the clearest case is provided
by the Ruijsenaars-Schneider quadratic Poisson structure involving two matrices~\cite{RS,Su}. One is then able~\cite{Su,ANR,ArChFr}
to identify the consistency equations as a classical limit of the Yang-Baxter equations for 
a dynamical reflection algebra. They include in particular the quantum Gervais Neveu Felder
cubic equation (symbolically written $RR^hR = R^h R R^h$) for e.g. Interaction Round a Face (IRF)-type quantum $R$-matrices, 
and its associated adjoint equation $RSS^h = SS^hR$.
It is interesting to remark that the first mentioned example, i.e.
the scalar Calogero Moser $r$-matrix, obeys in fact a consistent combination of these two equations.

Even in such cases when the $r$-matrix or matrices depend on Poisson-commuting
dynamical variables, the precise derivation of the associated classical Yang Baxter equation 
is thus a subtle procedure requiring
the explicit knowledge of the Lax matrix and $r$-matrix. Obtaining the 
specific form associated with the GNF equation is not a priori
guaranteed.

By contrast there are  fewer results regarding $r$-matrices depending on non-Poisson
commuting sets of dynamical variables (hereafter loosely denoted ``$p,q$ dependent
$r$-matrices''). Indeed few examples are known, even less explicitely so, 
no associated classical ``dynamical''
Yang Baxter equation has been derived, let alone any quantization thereof.
Note at this point that we expect any such quantization to have very unusual
features compared to the ``standard'' dynamical case, involving as it should
\textit{non-commutative} deformation parameters $p$, $q$ (instead of the
sole $c$-number parameters $q$) inside the generalized $R$ matrix.

The study of such $r$-matrices is thus clearly a very challenging and
promising issue in the domain
of integrable systems and this paper will present an attempt at
expanding the current state of knowledge about them by constructing
new examples of Lax matrices and their associated $r$-matrices.

To position the problem, let us first recall a few known examples where
the $r$-matrix exhibits these dependance: the complex sine Gordon
case~\cite{maillet2} (explicit pair $r,s$ but no decoupled Yang Baxter equation); 
the elliptic Calogero Moser model~\cite{BS1,BS2} in its spectral parameter-free formulation~\cite{OP} (nonexplicit);
the $BC_n$ Ruijsenaars Schneider Lax matrix~\cite{AR} (nonexplicit); and possibly
the recently proposed $2$x$2$ Lax matrix associated with special solutions of the Camassa Holm equation~\cite{HQ}.

The simplest example (albeit nonexplicit yet) is provided by the 
Ragnisco-Bruschi~\cite{BR} construction of the $r$-matrix for the Lax matrix yielding the Calogero-Fran\c{c}oise (CF) Hamiltonian~\cite{CF}.
The Lax matrix reads:
\be \label{LaxCalFran}
{\bf L} = \sum_{i,j} L_{ij}{\bf e_{ij}}\; ;\; L_{ij} ={\sqrt{p_i p_j}}\,A(q_i-q_j)\; ;\; 
{\rm{with}} \; A(q)=
\rm{cosh}(\nu\,q/2) + \rho \, \rm{sinh}(\nu \,\vert q/2 \vert)
\ee

The CF Hamiltonian is $Tr({\bf L}^2)$. This dynamical system may be viewed as a generalization of both
Camassa-Holm peakon integrable system (where $A(z) = exp (\nu |z|)$), describing the dynamics of specific singular ``peaked'' solutions to the famous Camassa Holm shallow wave fluid equation~\cite{CH93,CHH,ACHM}, 
and the related Calogero system~\cite{Cal} (where $A(z) = cos(z)$).

We shall be classifying here the Lax matrices \`a la Calogero-Fran\c{c}oise
admitting an $r$-matrix.  We shall in this paper restrict ourselves
to Lax matrices where the diagonal potential $A_{ii}$ is a constant (we shall
comment upon this at the end) and $A$ is an even function.  

The problem will then be solved by first imposing and solving a necessary condition on $A$ derived from the vanishing of the Poisson bracket 
$\{Tr{\bf L}, Tr {\bf L}^2\}$ and $\{Tr{\bf L}^2,Tr{\bf L}^3\}$; then explicitely constructing the $r$-matrix will show that any such solution 
be indeed integrable. 
Some further comments and perspectives shall be added at the end. 
Not surprisingly the periodic peakon potential~\cite{BSS1} will be obtained as one solution.

\section{The  model}

We assume the following form for the Lax matrix:

\be \label{Lax}
{\bf L}=\sum_{i,j} \sqrt{p_i\,p_j}\, A_{i,j}\,{\bf e_{ij}}, \quad \rm{with}\;
\left\{\begin{array}{l}
A_{i,i}=a_i \; \rm{some\; complex\;constant}\\ 
A_{i,j}=A(q_i-q_j) \;\rm{for}\; i\ne j 
\end{array}\right.
\ee
As usual ${\bf e_{ij}}$ denotes the $n \times n$ matrix representation of $gl(n)$.
We require that $A$ be an even complex function of a real variable, with isolated singularities (more precisely separated from
any other such point by a distance strictly bounded from below), at least $D_1$ around regular points and such that $A^2$ be a piecewise $D_2$ function.

\subsection{Necessary conditions}

Starting from the formulae for the first three Hamiltonians defined from ${\bf L}$:
\be \label{Ham}
&\ds  H_1=Tr {\bf L}=\sum_i p_i \, A_{i,i} =\sum_i p_i \, a_i \nn \\ 
&\ds  H_2=Tr {\bf L}^2=\sum_{i,j} p_i\,p_j\, A_{i,j} \, A_{j,i}=\sum_i p_i^2 \, a_i^2\;+ \sum_{i\ne j} p_i\,p_j\, A^2(q_i-q_j)
\\ 
&\ds  H_3=Tr {\bf L}^3=\sum_{i,j,k} p_i\,p_j\,p_k \, A_{i,j} \, A_{j,k}\, A_{k,i}=\nn \\
&\ds \sum_i p_i^3 \, a_i^3\;+ 3\,\sum_{i\ne j} p_i^2\,p_j\, a_i\, A^2(q_i-q_j)\,+ \sum_{i\ne j \ne k} p_i\,p_j\,p_k \, A(q_i-q_j \, A(q_j-q_k)\, A(q_k-q_i)\nn
\ee

we impose simultaneous vanishing of all Poisson brackets between them as a necessary condition for integrability.

With the convention $\{p_i,q_j\}=\d_{i,j}$, one gets
$$\{H_1,H_2\}= 2\, \sum_{i\ne j} p_i\,p_j\,(a_i-a_j)\, A(q_i-q_j)\, A'(q_i-q_j)$$
$H_1$ and $H_2$ are in involution if and only if either the potential is a piecewise constant function (a trivial solution, so we will solve excluding this case) or the diagonal part of the potential is uniform: $a_i=a$.
Setting all $a_i$'s to $a$ the Poisson bracket of $H_2$ and $H_3$, after some rearrangements, reduces to
$$\{H_2,H_3\}= 6\, \sum_{i,j,k,l} p_i\,p_j\,p_k \,p_{l} \,
  A_{k,l}\, (A_{i,l} \, A_{j,k}-A_{j,l} \, A_{i,k})\, 
(A_{l,i} \, A'_{l,j}-A_{l,j}\,A'_{l,i} )$$
with $A'_{i,j}=A'(q_i-q_j)$ for $i\ne j$ and $A'_{i,i}=0$ .
In particular the $p^2_a\,p_b\,p_c$ coefficient is a multiple of:
$\ds(A_{b,a}\,A_{a,c}-a\,A_{b,c})\, 
\left[(A_{a,b}^2-A_{a,c}^2) \, A'_{b,c} +(A_{a,b}\,A'_{a,b}+A_{a,c}\,A'_{a,c})\,A_{b,c}
-a\,(A_{a,b}\,A'_{a,c}+A_{a,c}\,A'_{a,b})\right]$.

Let us denote the differences in positions in terms of  $x=q_a-q_b$ and $y=q_c-q_a$; 
the commutation of $H_2$ and $H_3$  yields:
\be 
&A(x)\,A(y)=a\,A(y+x) \label{cond1}\\
&\rm{or}\nn\\
&(A^2(x)\!-\!A^2(y)) \, A'(y\!+\!x) +(AA'(y)\! -\! AA'(x))\,A(y\!+\!x)
=a\,(A(x)\,A'(y)\!-\!A(y)\,A'(x))\label{cond2}
\ee
Note that, if for $x$ and $y$ in the vicinity of some $x_o$ and $y_o$ equation (\ref{cond1}) is satisfied, 
then $A'(x)\,A(y)=A(x)\,A'(y)=a\,A'(y+x)$ yielding also
$A^2(x)\, A'(y\!+\!x)= AA'(x)\,A(y\!+\!x)$ and $A^2(y) \, A'(y\!+\!x)= AA'(y)\,A(y\!+\!x) $.
So equation (\ref{cond2}) has to be satisfied in the full plane $x,y$ except perhaps in some isolated points (necessarily singular for $A$).

Introducing the variables $s=y+x$ and $d=y-x$, as well as the three functions

$D(s,d)=A^2(y)\!-\!A^2(x)$, $F(s,d)=2\,A(y)\,A(x)$ and $\ds G=\pdi{d}F$,
 equation (\ref{cond2}) now reads:
\be  \label{cond}
\pdi{s}D(s,d)\,A(s)-D(s,d)\, A'(s)
=a\,G(s,d)
\ee
In the vicinity of a regular point, the derivation with respect to $d$ also yields:
\be  \label{condp}
\pdi{d}\pdi{s}D(s,d)\,A(s)-\pdi{d}D(s,d)\, A'(s)
=a\,\pdi{d}G(s,d)
\ee
Eliminating $A'(s)$ from equations (\ref{cond}) and (\ref{condp}) and defining $\ds \D=D\,\pdi{d}\pdi{s}D-\pdi{d}D\,\pdi{s}D$, one gets:
\be  \label{cos}
\D(s,d)\,A(s)=a\,(D(s,d)\,\pdi{d}G(s,d)-\pdi{d}D(s,d)\,G(s,d))
\ee 

Note that  $D(s,d)=D(d,s)$ directly from its definition since the exchange of $s$ and $d$ corresponds to the transformation of $x$ into $-x$ and $A$ is an even function. Hence $\D$ is also a symmetric function and one gets from (\ref{cos}):
\be  \label{cod}
\D(s,d)\,A(d)=a\,(D(s,d)\,\pdi{s}H(s,d)-\pdi{s}D(s,d)\,H(s,d))
\ee
with $\ds H=\pdi{s}F$ (note that F is also symmetric in $s$ and $d$).

Actually, by direct computation, one obtains that the two rhs of equations (\ref{cos}) and (\ref{cod}) are equal.  Let us denote them by $a\,K$ with $K$ the following symmetric function:
\be \label{K}
K(s,d)= D\,\pdi{d}\pdi{d}F-\pdi{d}D\,\pdi{d}F=
D\,\pdi{s}\pdi{s}F-\pdi{s}D\,\pdi{s}F
\ee
Since we have excluded from the begining any piecewise constant potential, we obtain $\D=0$ as a necessary condition:
\be  \label{ness}
\D=D\,\pdi{d}\pdi{s}D-\pdi{d}D\,\pdi{s}D=0
\ee
This very equation can be solved on any regular domain of the plane, so we get {\it piecewisely}: 
\be  \label{solD}
D(s,d)=\a(s)\,\b(d)
\ee
with $\a$ and $\b$ two $D_1$-functions on the interval of $s$, respectively $d$, where they build $D$. 
Note that if we change the domain of the plane where we solve the equation the functions $\a$ and $\b$ are not necessarily the same functions as in the first domain, even if they happen to be defined at some common value of $s$ (or $d$).

Remark that since $D(s,d)=A^2(y)-A^2(x)$ and since $A^2$ has been supposed piecewisely $D_2$, $\a$ and $\b$ are actually $D_2$, one thus gets {\it piecewisely}:
\be  \label{eqalphbet}
\frac{1}{4}\,\left((A^2)''(y)-(A^2)''(x)\right)=\a''(s)\,\b(d)=\a(s)\,\b''(d)
\ee
So for any  pair of functions $\a$ and $\b$, there exists some complex constant $\nu$ such that
\be  \label{solalphbet}
\a''(s)=\nu^2 \, \a(s)\;, \; \b''(d)=\nu^2 \, \b(d)\; {\rm and} \; (A^2)''(y)-4\,\nu^2 \,A^2(y)=(A^2)''(x)-4\,\nu^2 \,A^2(x)
\ee
Note that the last equality holds for any couple of regular points.
One finally obtains the following form for the square of the potential:
\be \label{A2form}
A^2(x)=\g+\l_x\,e^{2\,\nu_x\,x}+\mu_x\,e^{-2\,\nu_x\,x}
\ee
with $\g$ some complex constant and $\nu_x$, $\l_x$ and $\mu_x$ three piecewise constant functions (to discriminate between $\l$ and $\mu$ we will assume $Re(\nu_x)\ge0$).
Reinserting this form into (\ref{ness}) yields:
\be \label{plugDelt}
(\nu^2_x-\nu^2_y)\,(\l_x\,e^{2\,\nu_x\,x}+\mu_x\,e^{-2\,\nu_x\,x})\,
(\l_y\,e^{2\,\nu_y\,y}+\mu_y\,e^{-2\,\nu_y\,y})
+ 4\,(\nu^2_y\,\l_y\,\mu_y-\nu^2_x\,\l_x\,\mu_x)=0
\ee
One obtains that $\nu_x$ is actually a complex constant, i.e. $\nu_x = \nu_y $,
that we will accordingly denote by $\nu$ in the following , we can assume that it is not zero since we have already discarded the piecewise constant potential.
More interestingly the product $\l_x\,\mu_x$ is also constant and we shall now call it $\r^2$ (up to a normalization factor).

One then easily shows that $\r=0$ is essentially not compatible with the assumption of an even potential and leads to $A$ constant. In the following we will thus assume $\r\ne0$.

Let us now introduce another piecewise constant function $\t_x$, parametrize $\l_x\,\mu_x=\frac{\r^2}{16}$ by
$\l_x=\frac{\r}{4}\,e^{-\nu\,\t_x}$ and $\mu_x=\frac{\r}{4}\,e^{\nu\,\t_x}$, and get:
\be \label{A2form2}
&A^2(x)=\r\,({\rm cosh}^2(\nu\,(x-\t_x/2))-{\rm cosh}^2(\nu\,z_o))\nn\\
& =\r\,{\rm sinh}(\nu\,(x+z_0-\t_x/2))\,{\rm sinh}(\nu\,(x-z_0-\t_x/2))
\ee
with $z_o$ a complex constant.

This expression of $A^2$ yields:
\be \label{Dform}
&D(s,d)=\r\,({\rm cosh}^2(\nu\,(y-\t_y/2))-{\rm cosh}^2(\nu\,(x-\t_x/2)))\nn\\
& =\r\,{\rm sinh}(\nu\,(s-(\t_y+\t_x)/2))
\,{\rm sinh}(\nu\,(d-(\t_y-\t_x)/2))
\ee
Using this expression in equation (\ref{cond}), one shows that:
$$\pdi{d}(A(x)\,A(y))={\cal A}(s)\,\nu\,{\rm sinh}(\nu\,(d-(\t_y-\t_x)/2))$$
with ${\cal A}(s)$ some function of $s$.
Hence there exists some function ${\cal B}$  of $s$ such that:
\be \label{AxAy}
A(x)\,A(y)={\cal A}(s)\,{\rm cosh}(\nu\,(d-(\t_y-\t_x)/2))+{\cal B}(s)
\ee
From (\ref{A2form2}), one also gets:
\be \label{AxAy2}
&A^2(x)\,A^2(y)=\r^2
\,{\rm sinh}(\nu\,(x+z_0-\t_x/2))\,{\rm sinh}(\nu\,(y+z_0-\t_y/2))\nn\\
&\,{\rm sinh}(\nu\,(x-z_0-\t_x/2))\,{\rm sinh}(\nu\,(y-z_0-\t_y/2))\nn\\
&=\r^2/4\,({\rm cosh}(\nu\,(s+2\,z_o-(\t_y+\t_x)/2))-{\rm cosh}(\nu\,(d-(\t_y-\t_x)/2)))\nn\\
&({\rm cosh}(\nu\,(s-2\,z_o-(\t_y+\t_x)/2))-{\rm cosh}(\nu\,(d-(\t_y-\t_x)/2)))
\ee
The compatibility between (\ref{AxAy}) and (\ref{AxAy2}) imposes:
$${\rm cosh}(\nu\,(s+2\,z_o-(\t_y+\t_x)/2))={\rm cosh}(\nu\,(s-2\,z_o-(\t_y+\t_x)/2))$$
or equivalently ${\rm sinh}(\nu\,2\,z_o)=0$ restricting the constant  ${\rm cosh}^2(\nu\,z_o)$ to be  either $0$ or $1$.
Up to a shift on $\t_x$, we have obtained that the potential is of the following form:
\be \label{Aform}
A(x)=A_o\,{\rm cosh}(\nu\,(x-\t_x/2))
\ee
with $A_o$ a complex constant.
Equation (\ref{cond}) now reduces to an equation on the piecewise constant function $\t_x$ (defined modulo $4\,i\,\pi/\nu$):
\be \label{condtheat}
A_o\,{\rm cosh}(\nu\,(\t_x+\t_y-\t_{x+y})/2))=a
\ee
The only other condition is the parity of $A$, imposing $\t_{-x}=-\t_x$.
In particular this is true for $x=0$ where $A$ is thus continuous, but generically not differentiable.

So either $\t_x=\t_{0_+}=-\t_{-x}$ for all positive $x$, or it admits a first strictly positive discontinuity in $x_1$, and reads  $\t_x=(2\,n_x+1)\,\t_{0_+}$, with $n_x$ the integer quotient of $x$ divided by $x_1$ ($x=n_x\,x_1+r_x,\; r_x \,\in \,[0, x_1[$ and $n_x \,\in\,\Z$).
\\\\
Finally, we have established the following proposition:
\\\\
{\bf Proposition 1}
\\
A Lax matrix of type (\ref{Lax}) which defines a completly integrable system can at most depend on three complex parameters ($A_o, \nu$ and $\t_{0^+}$) and possibly a strictly positive real one ($x_1$) It has to be of the following form:
\be \label{LaxSol}
&\ds {\bf L}=\sum_{i,j} \sqrt{p_i\,p_j}\, A_o\,{\rm cosh}(\nu\,(q_i-q_j- \frac{_1}{^2}\,\t_{(q_i-q_j)}))\,{\bf e_{ij}}, \\
&{\rm with} \; {\rm either}\; \t_x=sign(x)\,\t_{0_+}\quad\rm{or}\quad
\left\{\begin{array}{l}
\t_x=(2\,n_x+1)\,\t_{0_+}\; \rm{for} \; x=n_x\,x_1+r_x,\\\
r_x \,\in \,[ 0, x_1 [, \; \rm{and} \;n_x \,\in\,\Z
\end{array}\right.\nn
\ee
\\\\
The first set of solutions with $\t_x=sign(x)\,\t_{0_+}$ corresponds 
exactly to the Calogero-Fran\c{c}oise Lax matrix (\ref{LaxCalFran}).
Note that within the new set of solutions, if one put $x_1=\t_{0_+}$, one gets a non trivial continuous  periodic potential.

\subsection{The $r$-matrix}
We will now establish Liouville-integrability for the Hamiltonians associated with the Lax matrix (\ref{LaxSol}) by explicitely constructing the $r$-matrix.

Following closely the $r$-matrix derivation in~\cite{BR} it is easily seen that
it hangs upon the following technical lemma:
\\\\
{\bf Lemma}
\\
If there exist two odd functions $B$ and $C$, such that the even function $A$ obey the consistency equation (labeled $(2.10)$ in ~\cite{BR}) where again $s = x+y$ :
\be \label{Afunc}
\pdi{s}(A(x)\,A(y))=B(s)+A(s)\,(C(x)+C(y))
\ee
the Lax matrix $\ds {\bf L}=\sum_{i,j} \sqrt{p_i\,p_j}\, A(q_i-q_j)\,{\bf e_{ij}}$ has a linear $r$-matrix Poisson structure: $$\ds \{{\bf L}_1,{\bf L}_2\}~=~[{\bf r}_{12},{\bf L}_1]-[{\bf r}_{21},{\bf L}_2]\quad \mbox{with the standard notations}\; {\bf L}_1\equiv {\bf L}\otimes {\bf{1}} \; \mbox{and} \; {\bf L}_2\equiv{\bf{1}}\otimes {\bf L}$$
and
$\ds {\bf r}_{12} \equiv \sum_{i,j,k,l} \, r_{ijkl} \, {\bf e_{ij}} \otimes {\bf e_{kl}} \,\, ; \,\,
{\bf r}_{21}= \sum_{i,j,k,l} \, r_{ijkl} \, {\bf e_{kl}} \otimes {\bf e_{ij}}
$ given by:
\be \label{matr}
\ds{\bf r}_{12}=\frac{_1}{^2}\,\sum_{i\ne j} C(q_i-q_j)\,{\bf e_{ij}}\otimes({\bf e_{ij}}+{\bf e_{ji}})+ \,\frac{_1}{^2}\,\sum_{i, j, k} S_{j k} \,({\bf e_{ij}}-{\bf e_{ji}})\,\otimes\,({\bf e_{ik}}+{\bf e_{ki}})
\ee
$S_{j k}$ are the elements of any symmetric matrix ${\bf S}$ solving the algebraic equation:
\be \label{matS}
[{\bf S}, {\bf L}]_{i j}=\sqrt{p_i\,p_j}\, B(q_i-q_j)
\ee
\\\\
{\bf Proof}: 
Proof can be obtained from direct verification, once it is established that the Poisson brackets of ${\bf L}$ with itself are given uniquely in terms of the function $\pdi{s}(A(x)\,A(y))$ .
\\\\
{\bf Remark 1}

Existence of a solution ${\bf S}$ to the algebraic equation (\ref{matS}) is obtained by rewriting the defining
equation in an orthogonal basis for the symmetric matrix ${\bf L}$ when ${\bf L}$ is real. If ${\bf L}$
is complex symmetric, one needs to assume separately that it can be diagonalized (which
is true if ${\bf L}$ is in general position). In both cases one also needs to assume that
the eigenvalues are non degenerate. Note that the issue of degenerate eigenvalues for
a Lax matrix is in any case a delicate one and needs a case-by-case treatment.
\\\\
{\bf Remark 2}

${\bf S}$ is in any case not unique, since in particular
 any polynomial in ${\bf L}$ yields a symmetric matrix commuting
with ${\bf L}$ itself.

We are currently working on an explicit formulation of solutions to equation (\ref{matS}).
\\\\
We now state the key result of this section:
\\\\
{\bf Proposition 2}
\\
The function $A(x)$ in (\ref{Aform}) obeys the $r$-matrix consistency equation (\ref{Afunc})
with the functions:
\be\label{foncB}
& B(x) = \frac{1}{2} (A'(x)\,A(0)-\frac{\t_x}{\t_{0_+}} \,A(x)\,A'(0))\\
&= \frac{\nu\,A_o^2}{2} ({\rm sinh}(\nu\,(x-\t_x/2))\,{\rm cosh}(\nu\,\t_{0_+}/2)
+\frac{\t_x}{\t_{0_+}} \,{\rm cosh}(\nu\,(x-\t_x/2))\,{\rm sinh}(\nu\,\t_{0_+}/2))\nn
\ee
\be\label{foncC} {\rm and}\quad C(x)=\frac{1}{2} \frac{\t_x}{\t_{0_+}} \,A'(0)
=-\frac{\nu\,A_o}{2} \frac{\t_x}{\t_{0_+}} \,{\rm sinh}(\nu\,\t_{0_+}/2)\ee

Oddness of functions $B$ and $C$ follows from $\t_{-x} = -\t_{x}$. The proof 
proceeds by direct checking
of the ``addition'' formula (\ref{Afunc}), paying due attention to the
consistency of this addition for the step function $\theta_x$.

Hence all necessary potentials $A$ yield a Lax matrix which exhibits an $r$-matrix structure of
the form obtained by Ragnisco-Bruschi, albeit with periodic or more
generally pseudoperiodic coefficient functions $B$ and $C$. We have thus completely
classified the integrable classical Lax operators of the Calogero-Fran\c{c}oise form.

\section{Remarks and perspectives}

1. We have obtained a new set of pseudoperiodic potentials as a result
of our systematic search for even Calogero-Fran\c{c}oise type potentials.
It would be interesting to know how to formulate alternatively {\cal algebraic}
integrability properties (extension of the Abel maps and associated hyperelliptic
curves~\cite{BSS1}) for such pseudoperiodic potentials. In principle
the conserved Hamiltonians should be moduli of the associated curve.
The possible connection with the pseudoperiodic {\it solutions} derived
in e.g.~\cite{ACHM} may also be interesting.

Of course among these potentials one recovers the fully periodic 
peakon potential derived by Beals et al~\cite{BSS1}. Here we prove its $r$-matrix
(hence Liouville) integrability whereas~\cite{BSS1} had proven algebraic integrability
in the above sense. 

2. Interpretation of the pseudoperiodic ``peakons'' in the frame of Camassa Holm
equation may be interesting, possibly as solutions for a system in a finite
volume with some specific boundary conditions.

3. It is easily checked that a folding procedure exists which consistently yields
new integrable systems from these initial pseudoperiodic ones. Starting from the
$N= 2n (+1)$-body system with positions labeled now from $-n$ to $+n$
(without/with 0) one proves immediately that the folding constraints
$p_i = p_-i$, $q_i = q_-i$ are invariant by evolution under all even (resp. odd) degree 
Hamiltonians. Hence the reduced dynamical system itself has $n$ invariant Poisson
commuting Hamiltonians under the reduced Poisson bracket. Its Lax matrix takes the form: 
$\ds {\bf L} = \sum_{i,j = 1 (0)}^{n} L_{ij}\,({\bf e_{ij} + e_{-i,j} + e_{i,-j} + e_{-i,-j}})$ with:

\be \label{LaxCalFranFol}
L_{ij} ={\sqrt{p_i p_j}}\,({\rm{cosh}}{\nu  (q_i -q_j)\over 2}~+~\rho \, {\rm{sinh}}{\nu \vert q_i -q_j \vert
 \over 2})
\ee

Here one first (rather trivial) possibility of
extension of the strict Calogero Fran\c{c}oise framework is made explicit.

4. In order to proceed towards a deduction of the form
for the Yang-Baxter type equation one key point now is to get an explicit form for the $r$-matrix
by solving the matrix equation for ${\bf S}$. Partial results are now available and
we hope to soon have the general form of this $r$-matrix.

\vspace{.75cm}
{\bf Acknowledgements}
\vspace{.5cm}

We wish to thank Eric Ragoucy for his interest and his collaboration in unraveling the algebraic equation (\ref{matS}). J.A. thanks LAPTH Annecy for their kind hospitality.

\end{document}